\documentclass[a4paper]{iopart}
\usepackage[cp850]{inputenc}
\usepackage{mathptmx}
\usepackage{geometry}
\usepackage{graphicx}
\usepackage{multicol}
\usepackage[
   colorlinks,        
   linkcolor=blue,   
   filecolor=blue,   
   citecolor=blue   
]{hyperref}
\usepackage{textcomp}
\begin{document}

\title[La$_{0.7}$Ce$_{0.3}$MnO$_3$ thin 
films under photoexcitation]
{Conductivity and magnetoresistance of La$_{0.7}$Ce$_{0.3}$MnO$_3$ 
thin films under photoexcitation}

\author{A Thiessen$^1$, E Beyreuther$^1$,
R Werner$^2$, R Kleiner$^2$, D~Koelle$^2$ and L~M~Eng$^1$}

\address{$^1$Institut f\"ur Angewandte Photophysik, 
Technische Universit\"at Dresden, D-01062 Dresden, Germany \\
$^2$Physikalisches Institut and Center for Collective 
Quantum Phenomena in LISA$^+$,
Universit\"at T\"ubingen, Auf der Morgenstelle 14, D-72076 T\"ubingen, Germany}

\eads{\mailto{andreas.thiessen@iapp.de}, \mailto{elke.beyreuther@iapp.de}}

\begin{abstract}
La$_{0.7}$Ce$_{0.3}$MnO$_3$ thin films of different thicknesses, degrees 
of CeO$_2$-phase segregation and oxygen deficiency, grown on SrTiO$_3$ single 
crystal substrates, were comparatively investigated with respect to both
their spectral and temperature-dependent photoconductivity (PC) and 
their magnetoresistance (MR) behaviour under photoexcitation.
While as-grown films were insensitive to optical excitation, oxygen 
reduction appeared to be an effective way to decrease the film resistance, 
but the film thickness was found to play a minor role. However, from the 
evaluation of the spectral behaviour of the PC and the comparison of the 
MR of the LCeMO/substrate-samples with a bare substrate under illumination 
we find that the photoconductivity data reflects not only 
contributions from (i) photogenerated charge carriers in the film and (ii)
carriers injected from the photoconductive substrate (as 
concluded from earlier works), but also (iii) a decisive parallel 
photoconduction in the SrTiO$_3$ substrate.
Furthermore -- also by analyzing the MR characteristics -- 
the unexpected occurence of a strong electroresistive effect 
in the sample with the highest degree of CeO$_2$ segregation and oxygen 
deficiency could be attributed to the electroresistance of the SrTiO$_3$
substrate as well. The results suggest a critical reconsideration and 
possibly a reinterpretation of several previous photoconductivity and 
electroresistance investigations of manganite thin films on SrTiO$_3$. 
\end{abstract}


\section{Introduction}
\label{sec_Introduction}

Doped rare-earth manganites, also referred to as mixed-valence or 
colossal-magnetoresistive (CMR) manganites, 
have attracted a flurry of interest for decades \cite{ram97,coe99,hag03,dor06}. 
They are primarily of academic interest within the study of 
fundamental concepts in solid state physics addressing issues such as 
their multifaceted electronic/magnetic phase diagrams, their strong coupling 
between lattice, charge, spin, or orbital degrees of freedom, their sensitivity 
towards external fields, the high spin polarization in some compounds, or
the occurrence of electronic phase separation in chemically homogeneous 
samples. While the technological value of 
bulk crystals seems to be limited, e.g. due to the low magnetic Curie temperatures, 
thin films of selected compositions have the potential to play a role within 
tunable heterostructures in future all-oxide/all-perovskite electronic devices.

The present work focuses on two of the younger aspects of
manganite-compound research: 
the possibility of tetravalent-ion substitution in order to achieve effective 
electron-doping, 
e.g. for fabricating all-manganite diodes, and the behaviour under external optical 
excitation.

For the last two decades, \emph{tetravalent-ion substituted} rare-earth 
manganites (with La$_{1-x}$Ce$_x$MnO$_3$ 
being the most prominent among them) have been discussed intensively as a 
potential electron-doped counterpart to the well-known divalent-ion 
substituted and thus hole-doped manganites. A summary of the debate, focusing 
on possible (thin film) preparation routes and the proof of an effective 
electron-doping, i.e., a mixed Mn$^{2+/3+}$ valence, has been 
given in previous works \cite{bey06,wer09}. For instance, as a number of 
investigations have shown, as-prepared
La$_{0.7}$Ce$_{0.3}$MnO$_3$ (LCeMO) films do not show their nominal 
Mn valence of $+$2.7 but are
in fact hole-doped due to oxygen excess. On the other hand, 
post-deposition oxygen reduction decreases the Mn valence
towards the nominal value but also leads to a pronounced 
resistance increase and the loss of the manganite-typical metal-insulator
transition (MIT). Interestingly, a decisive conductivity increase 
and the recovery of the MIT can be induced by \emph{photoexcitation} \cite{bey09}. 
However, the exact origin of these effects remained hidden to date. 
In very recent 
x-ray photoemission measurements, exploring 
the Mn valence as a function of oxygen content 
and illumination intensity in variably thick LCeMO films on SrTiO$_3$, 
evidence for the decisive role of photogenerated-charge 
injection from the substrate to the films was reported, but an additional 
manganite-intrinsic photoconductivity could not be completely 
excluded \cite{thiessen_mn2+/mn3+_2014}. This is one of the starting points 
where the present work sets in. 

Compared to the number of investigations reflecting on the tunability 
of the manganites' conductivity 
and other physical properties by doping, magnetic or strain fields, studies of 
\emph{the effects of photonic excitations on manganites} are 
still a quite small research field. 
For a further discussion we define the 
photoresistance $PR$ as follows: 

\begin{equation}
PR:=\frac{R_{dark}-R_{illum}}{R_{illum}}=\frac{R_{dark}}{R_{illum}}-1,
\end{equation}

with $R_{dark}$ being the sample resistance in the dark, and $R_{illum}$ being 
the resistance under light excitation. Thus a decrease of the resistance under 
illumination results in a \emph{positive} PR, i.e., an appreciable
\emph{photoconductivity}. 

A survey of the previous literature is 
given in \cite{bey09, bey10}. In short, after the first discovery of 
photoconducticity in a mixed-valence manganite, namely in a 
Pr$_{0.7}$Ca$_{0.3}$MnO$_3$ single crystal \cite{kir97}, a number of 
single-crystalline as well as thin-film manganite compounds were studied 
under various photoexcitation conditions, and a plenty of effects were
reported. A general coherent microscopic picture of the photoresponse is 
still lacking to date. 
Roughly, the effects belong to one of the four following groups: 

\begin{itemize} 

\item photonic excitation of electrons from the valence to the conduction band 
\cite{gil00,cau01}, 
\item photoinduced demagnetization \cite{mat08,gao09}, 
\item disturbance of 
charge and orbital order in insulating charge ordered manganites by 
photoelectrons \cite{fie98,fie99,smo01,cha08}, 
and
\item injection of photogenerated charge 
carriers from the substrate into the thin-film channel 
\cite{bey09,bey10,kat00a,kat01}. 

\end{itemize}

The last of these four mechanisms 
naturally occurs in thin films only; but there it plays a major and possibly 
technologically interesting role. In La$_{0.7}$Sr$_{0.3}$MnO$_3$ films on 
SrTiO$_3$, which were illuminated 
with a broad-band white-light source and exhibited a \emph{negative} PR, Katsu 
et al. concluded an injection of optically generated electrons from the 
SrTiO$_3$ substrate into the hole-doped film followed by the recombination 
of both carrier types within the film leading to a resistance increase 
\cite{kat00a,kat01}. In our two previous 
works, where we observed \emph{positive} PR in 
La$_{0.7}$Ce$_{0.3}$MnO$_{3-\delta}$ \cite{bey09} and 
La$_{0.7}$Ca$_{0.3}$MnO$_{3-\delta}$ \cite{bey10}
films on SrTiO$_3$, we gave a similar interpretation. 
By the comparative evaluation of the wavelength dependence 
of the PR and the surface photovoltage we had concluded that at least part of 
the 
photogenerated carriers must stem from interband transitions in the 
substrate or from carrier excitation from interface states, the latter 
explaining the PR below the SrTiO$_3$ band gap energy, and not solely from the 
manganite film. However, these 
investigations were limited to a single film with a thickness of 10~nm 
only.

Thus, the present experiments were planned and conducted purposely 
to clarify the microscopic scenario of the 
phototransport in Ce-doped manganite thin films employing a broader 
set of films.

Extended resistance-vs.-temperature (\emph{R-T})
data of the films under different illumination conditions was acquired
in order to systematically analyse the influence of film thickness, oxygen content, 
and degree of phase separation on (i) the dark transport, which is 
reported in a separate paper \cite{thiessen_hopping_2014}, and (ii) 
the phototransport, which is discussed in section~\ref{sec_R_T}. 

Moreover, due to the well-known strong coupling of magnetic 
order and electronic transport in the manganites it appeared to be 
indispensable to study the magnetotransport under illumination for a further 
understanding of the microscopic nature of 
the observed photoconductivity. The corresponding 
magnetoresistance (MR) measurements 
and results for selected samples are presented in section~\ref{sec_MR_exp}.

Finally, the LCeMO film exhibiting the strongest disorder showed 
(apart from its photoconductivity) a rather unexpected electroresistive 
effect. There, magnetoresistance measurements were employed to understand 
this (side) effect as well.

\section{Experimental}


Three different LCeMO films were grown by pulsed laser deposition on SrTiO$_3$ 
(100) single crystal substrates. 
Details on growth and structural analysis can be found in 
reference~\cite{wer09}. Samples\footnote{Note that
sample labelling begins with B and not with A in order to keep 
the sample names of our previous works \cite{bey09,thiessen_mn2+/mn3+_2014}.} 
B and C were grown under the same oxygen 
partial pressure of p(O$_2$)$=$0.25~mbar but 
are different in thickness (30~nm vs. 100~nm), while samples 
C and D have the same thickness but D was grown under a decisively lower 
oxygen pressure, which led to microscopic segregation of CeO$_2$ clusters. Thus, 
the comparison of B and C provides 
information on the thickness dependence of the 
films' photoresponse, and contrasting C and D is supposed to give comparative 
insight to the PCs dependence on the degree of phase 
segregation.

To separate the influence of the oxygen stoichiometry from any of the 
above parameters, each sample was studied at three different states of 
oxygen reduction, which were prepared by heating in a low-pressure 
oxygen atmosphere (see table~\ref{tab_2}).

For reference measurements of the magnetoresistance, also 
a \emph{bare} SrTiO$_3$ (100) single crystal was reduced by applying the 
same procedure as for preparing the \emph{highly reduced} state 
of the thin-film samples.

\begin{table}
\caption{\label{tab_2}Overview of the heating parameters for the 
preparation of
differently oxygen-reduced LCeMO/SrTiO$_3$ samples. Note that each 
of the states was prepared for each of the three samples B, C, D. The 
\emph{as-prepared} state means that the sample has not been furtherly 
treated after the PLD growth.}
\begin{indented}
\lineup
\item[]\begin{tabular}{@{}lccc}
\br
state& T ($\symbol{23}$C) &p(O$_2$) (mbar)& t (h) \\
\mr
as-grown		&	--	&	--	&	--	\\
slightly reduced & 480  & 10$^{-6}$ & 1\\
highly reduced & 700  & 10$^{-8}$ & 2\\
\br
\end{tabular}
\end{indented}
\end{table}


The illumination setup consists of the monochromatized output of a Xe arc lamp, 
several collimating and focusing lenses, mirrors, and a control loop for 
adjusting the photon flux by moving a neutral density filter, 
mounted on a motorized 
translation stage. Details can be found in~\cite{bey09,bey11}.

The photoconductivity measurements were carried out in an optical 
liquid-nitrogen cryostat (Optistat DN by Oxford Instruments). The illuminated 
area on the sample was around 20~mm$^2$.
The magnetotransport measurements were performed in an optical 
liquid-helium cryostat with a superconducting magnet (Microstat MO by 
Oxford Instruments) producing a magnetic field $B$ perpendicular to the 
sample surface. Here, a sample area of 8~mm$^2$ was illuminated.
All resistance measurements were carried out in two-point geometry 
using an electrometer (Keithley 6517B), applying a measuring voltage of 
1~V, unless stated otherwise. Contacts were made with 
conducting silver paste, placed 3~mm apart.

\section{Results and Discussion}

\subsection{Photoconductivity}
\label{sec_R_T}


As a first result, \emph{no change} of 
the \emph{R-T} characteristics under illumination could be observed 
in the three \emph{as-prepared} films. 
This is in accordance with our former XPS study 
\cite{thiessen_mn2+/mn3+_2014}, where the same samples showed no change of 
the Mn valence under additional photoexcitation in the visible and 
near-UV range, as well.


\begin{figure}
\centering
\includegraphics[width=0.75\textwidth]{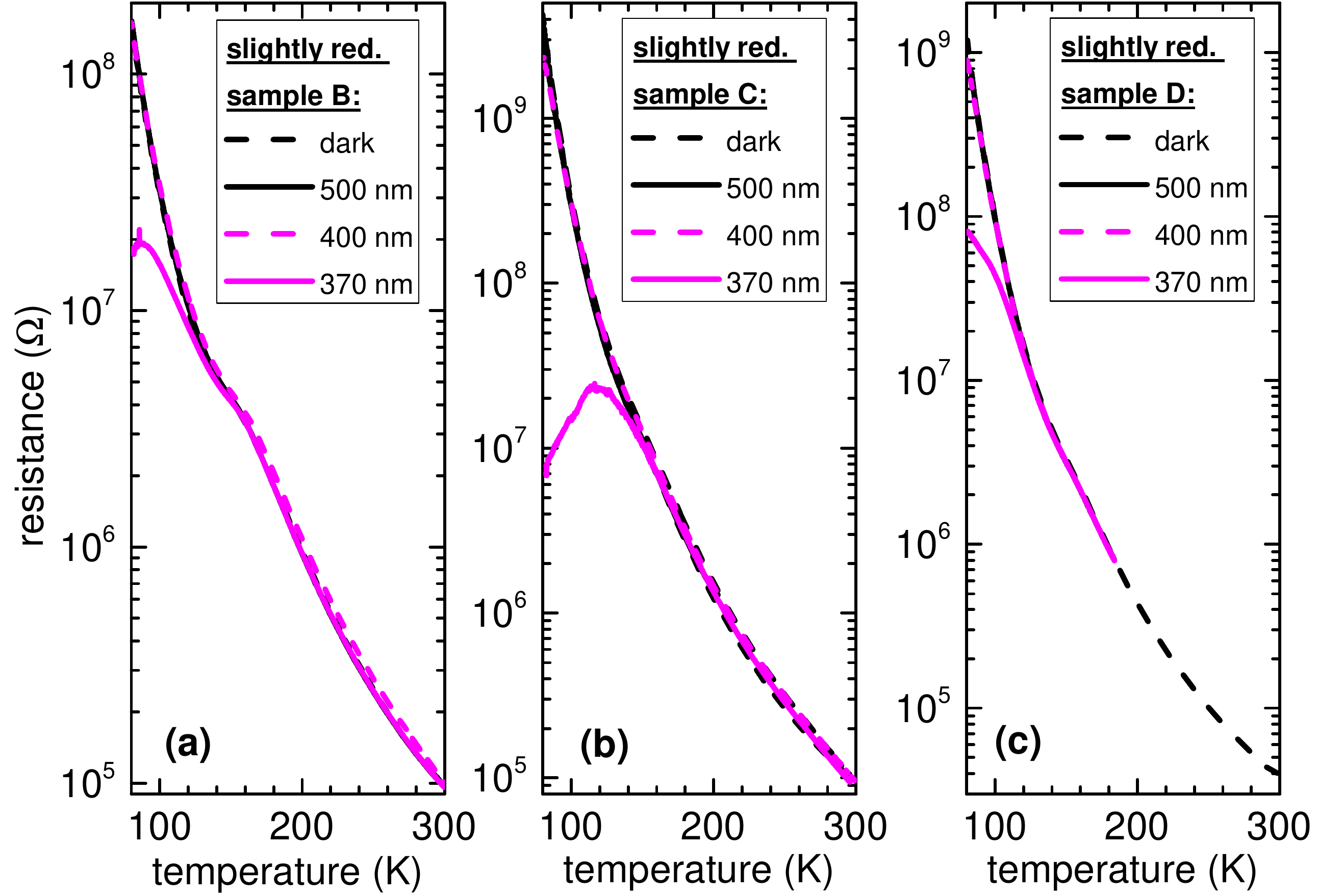}
\caption{\label{fig1} (a)-(c): \emph{Slightly reduced} LCeMO films B, C, and D:
Temperature dependence of the resistance in the dark and under illumination 
with three different wavelengths. The light intensities were 810~\textmu W/mm$^2$ 
at 370~nm, 750~\textmu W/mm$^2$ at 400~nm, and 600~\textmu W/mm$^2$ at 500~nm.}
\end{figure}

\begin{figure}
\centering
\includegraphics[width=0.75\textwidth]{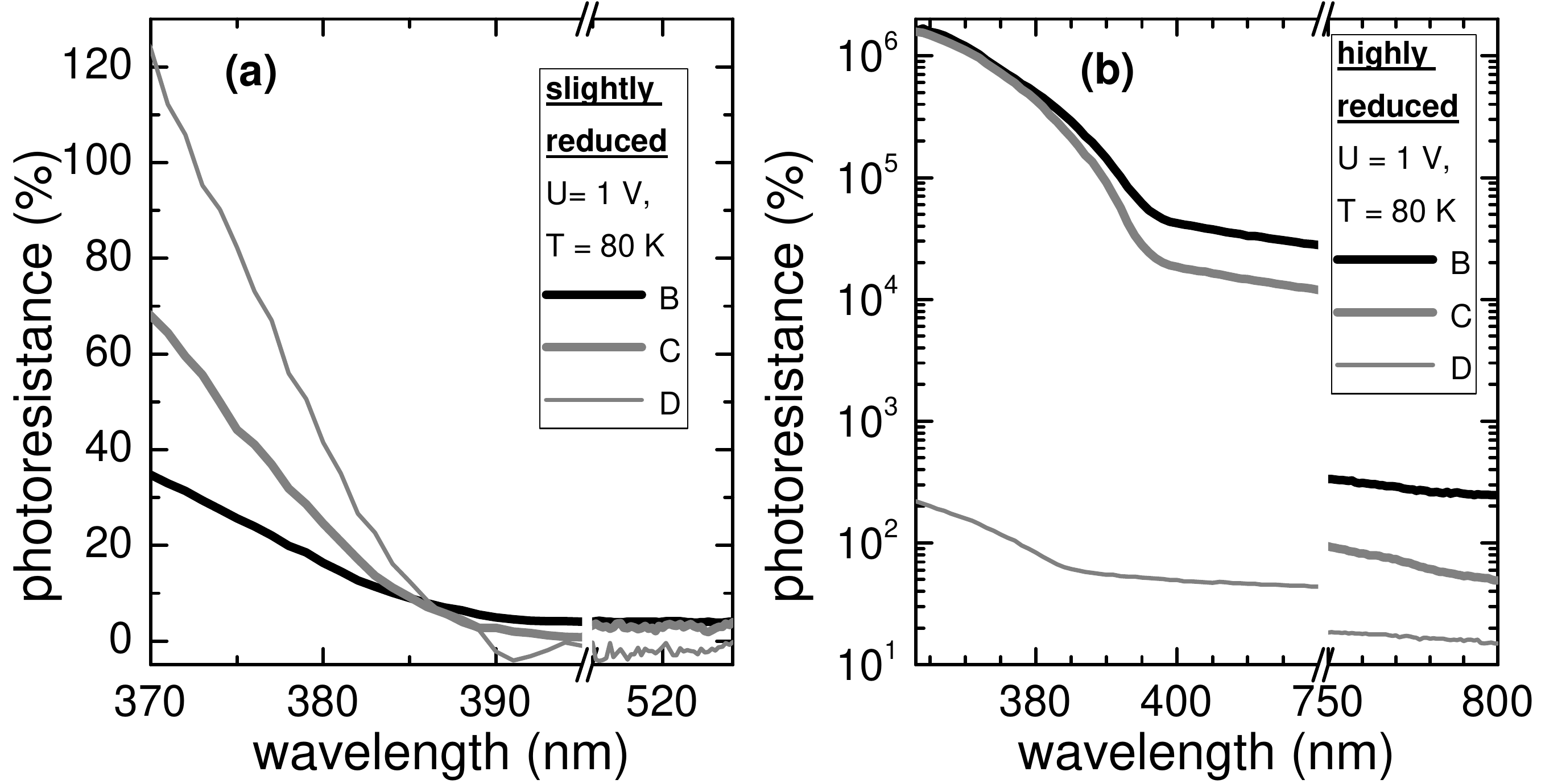}
\caption{\label{fig2} Photoresistance of (a) the \emph{slightly reduced} and 
(b) the \emph{highly reduced} LCeMO films at 
80~K, recorded with a constant photon flux of 6.8~\textmu W/mm$^2$ at 370~nm. Note 
that the data was recorded with a photon flux around two orders 
of magnitude lower than in the measurements of figs.~\ref{fig1} and~\ref{fig3}, since 
here the monochromator exit slit was reduced to 1~mm width (instead of 
10~mm) in order to achieve a higher energy resolution and thus a 
sharper photoresistance spectrum.}
\end{figure}

At second, let us consider the photoresponse of the \emph{slightly reduced} samples. 
Figure~\ref{fig1} depicts 
the \emph{R-T} curves in the dark and under illumination at three different wavelengths, 
i.e., 370~nm 
(above the SrTiO$_3$ bandgap $E_g=$~3.2~eV$\simeq$~387~nm), 
400~nm (slightly below $E_g$), 
and 500~nm (substantially below $E_g$). There is no resistance change 
under illumination at room temperature. A significant 
photoinduced decrease of the resistance 
is observed below 150~K and only for the 370-nm illumination. For sample C the 
MIT appears to be recovered. Figure~\ref{fig2}(a) contains the 
spectral photoresistance (PR) data at 80~K. Here, a noticeable increase of the 
PR is observable below 385$\ldots$390~nm (corresponding to $E_g$) 
for all three samples.


\begin{figure}
\centering
\includegraphics[width=0.75\textwidth]{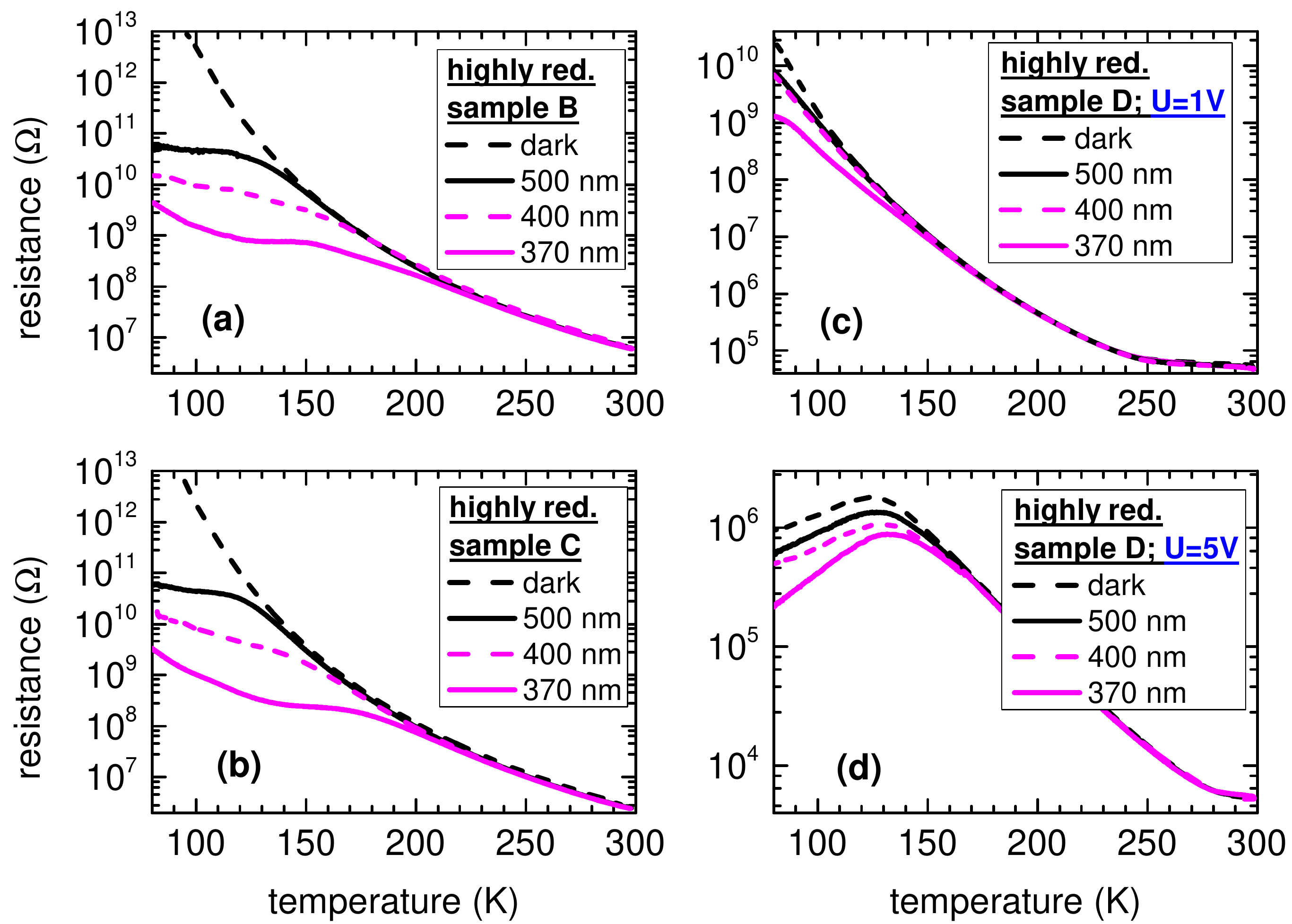}
\caption{\label{fig3} \emph{Highly reduced} LCeMO films: \emph{R-T} characteristics 
in the dark and under illumination (at the same light intensities as in 
figure~\ref{fig1}). (a), (b) Samples B and C at 1~V measurement voltage. 
(c), (d) Sample D at 1~V and 5~V measurement voltage.}
\end{figure}

Proceeding with the discussion of the \emph{highly reduced} LCeMO films, 
as expected, the dark resistances
are substantially higher than in the slightly reduced samples, 
see figure~\ref{fig3}. At a measuring voltage of 1~V, all samples exhibit 
an insulating behaviour in the whole investigated temperature range and no 
clear MIT. Similarly 
to the slightly reduced films, illumination leads to a significant decrease of the 
resistance at low temperatures, but makes no effect at 300~K. Different to the 
case of the slightly reduced films [cf. figure~\ref{fig2}(a)] 
which exhibit moderate PR values up to 120~\%, the PR at 80~K shows 
dramatically high values already for longer wavelengths, see 
figure~\ref{fig2}(b). Below 
395~nm (B, C) and 385~nm (D) the PR increases even more strongly with 
decreasing wavelength. Remarkably, the 
PR of the thinner (30~nm) film B is higher than the PR of the thicker (100~nm) film C.
Assuming an intrinsic charge carrier generation in the film, one would expect 
the opposite to happen, since the thicker film must absorb more light and thus exhibit the 
larger PR. If charge carrier injection from the substrate is the dominating process, 
the thinner film should, as indeed observed here, have the higher PR, since 
it passes more photons to the substrate. Thus, at a first glance, the current 
observation
points towards the charge injection scenario postulated before 
in ref.~\cite{bey09}. On the other hand, regarding the case of the 
slightly reduced films, the PR of the thinner film~B is \emph{lower} than the 
PR of the thicker films C and D [see fig.~\ref{fig2}(a), again]. 
It is conceivable, that both processes, intrinsic carrier generation in 
the LCeMO and injection of photogenerated carrier from the substrate 
coexist, but with different weight depending on the degree of oxygen 
reduction.

\begin{figure}
\centering
\includegraphics[width=0.75\textwidth]{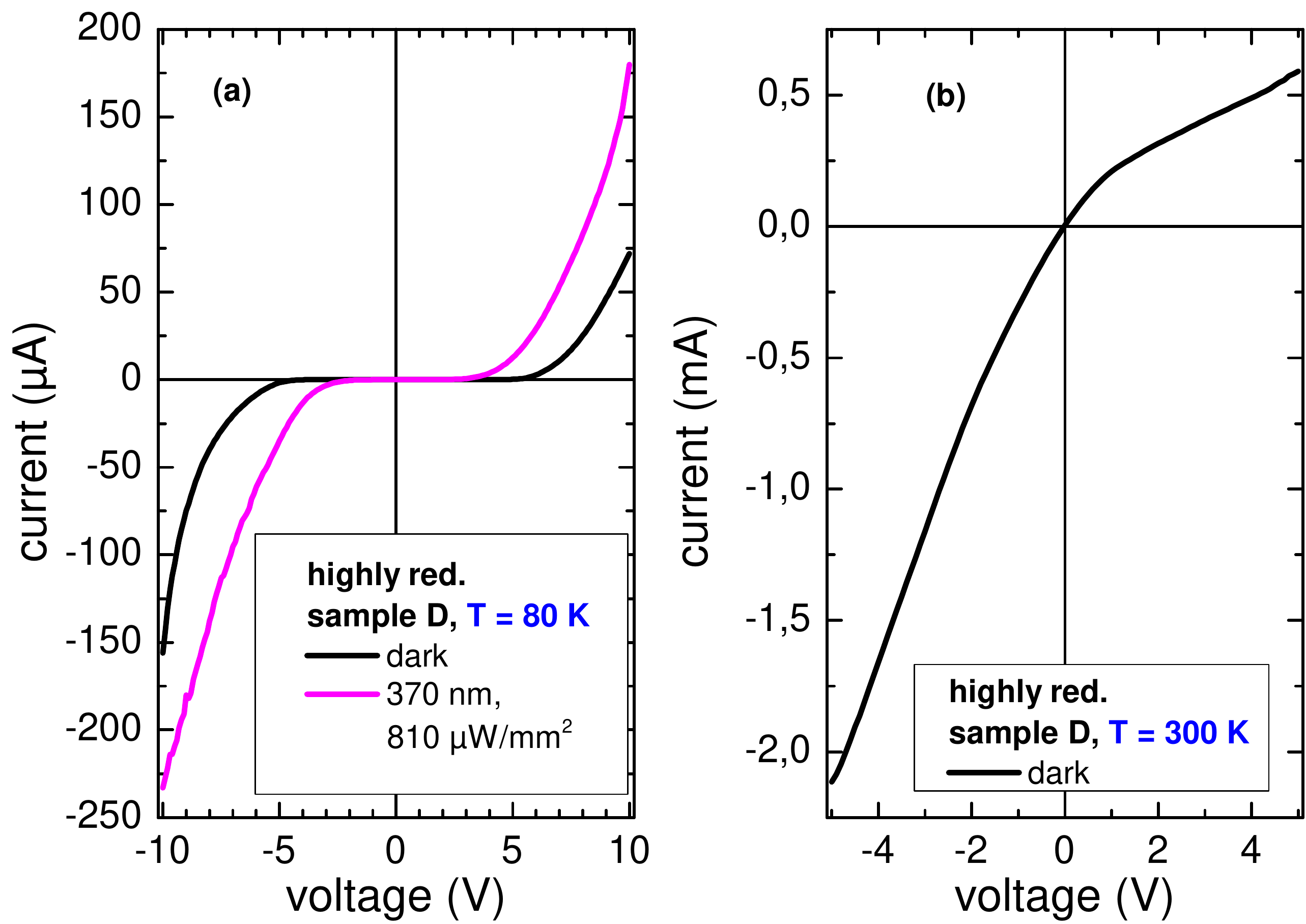}
\caption{\label{fig4} Current-voltage characteristics of the highly reduced 
sample D at 80~K (a) and at 300~K (b).}
\end{figure}

Interestingly, the resistance of film D strongly depends on the 
measuring voltage. As visible in figure~\ref{fig3}(d), the resistance at 5~V is 
at least one order of magnitude lower than at 1~V [figure~\ref{fig3}(c)] 
over the whole temperature range investigated here. Furthermore, at 5~V, an MIT 
appears at 135~K, whose transition temperature rises under illumination. Indeed, 
as shown in figure~\ref{fig4}, the current-voltage characteristics of the 
highly reduced sample D is highly nonlinear and asymmetric at 300~K as well 
as at 80~K, which points towards an enhanced inhomogeneity of the film. We note 
that the corresponding I-V curves of samples B and C (not shown here) 
exhibit a linear shape at 
300~K and nonlinear asymetric shapes at 80~K, but the latter with very low 
currents in the pA range, which limits their significance.

\subsection{Magnetoresistance}
\label{sec_MR_exp}

Due to the well-known strong interplay between charge transport and 
magnetic order in the mixed-valence manganites, we performed 
systematic measurements of the magnetoresistance as a function of 
magnetic field and temperature in order to achieve a profound understanding of the 
effect of illumination and measuring voltage on the charge transport in 
the \emph{highly reduced} LCeMO samples.
Thus, an extended temperature range between 5~K and 300~K was studied using a 
liquid-helium cryostat for the following selected samples:

\begin{itemize}

\item The \emph{highly reduced sample C} was measured in order to clarify the 
origin of the photoconductivity and the photoinduced MIT.

\item The \emph{highly reduced sample D} was studied in order to clarify the 
origin of the electrical-field-induced MIT. Since the interpretation of those 
results turned out to be complex and the electroresistance is more a 
side effect within the present study, we have exported the information 
and discussion for this sample into the supplement.

\item The \emph{as-prepared sample C} and a \emph{reduced bare 
SrTiO$_3$} single crystal were chosen for reference measurements.

\end{itemize}

Prior to presenting our experimental findings, we briefly recall the 
necessary background needed to evaluate the MR data:

In the literature on 
the CMR effect, the following definition of the $MR$ is 
frequently used:

\begin{equation}
\label{MR_definition}
	MR=\frac{R(B)-R(0)}{R(0)}=\frac{R(B)}{R(0)}-1   \quad  .
\end{equation} 

In mixed-valent manganites, the resistance at zero field $R(0)$ is 
larger than the resistance in an external magnetic field $R(B)$ (negative MR). 
Then the above definition gives values between -1 and 0.

A quantitative model describing the field dependence of the MR is given by 
Wagner et al. \cite{wag98}. Here, the following proportionalities 
are obtained:

\begin{equation}
\label{eq_Brillouin_2}
	MR(B)\propto \mathcal{B}_J^2[g \mu_B J B/(k_B T)] 
	\quad \mbox{for} \quad T>T_C \quad ,
\end{equation}
 and
\begin{equation}
\label{eq_Brillouin_1}
	MR(B)\propto \mathcal{B}_J[g \mu_B J B/(k_B T)] 
	\quad \mbox{for} \quad T<T_C \quad ,
\end{equation}

with $\mathcal{B}_J$ being the Brillouin function, $J$ the total 
angular momentum of the magnetic moments at the 
starting and the ending position of a hopping process, and $g$ the gyromagnetic 
factor with $g=2$. Furthermore, from geometrical considerations the 
following relation to 
calculate the radii of magnetic clusters $r_{MC}$ in the 
paramagnetic phase (which were also interpreted 
as magnetic polarons \cite{coe95}) can be derived:

\begin{equation}
\label{eq_mag_pol_size}
	r_{MC}=\left(\frac{3}{4\pi}\frac{J}{J(Mn^{3+})}\right)^{1/3}\cdot c \quad ,
\end{equation}

with $J(Mn^{3+})=2$ (valid for the Mn$^{3+}$ ion without magnetic 
interactions) and $c$ being the lattice constant. In this model, 
the ratio 
$J/J(Mn^{3+})$ is an estimate of the number of magnetically 
coupled Mn ions.

\subsubsection{As-prepared sample C}

\begin{figure}
\centering
\includegraphics[width=0.75\textwidth]{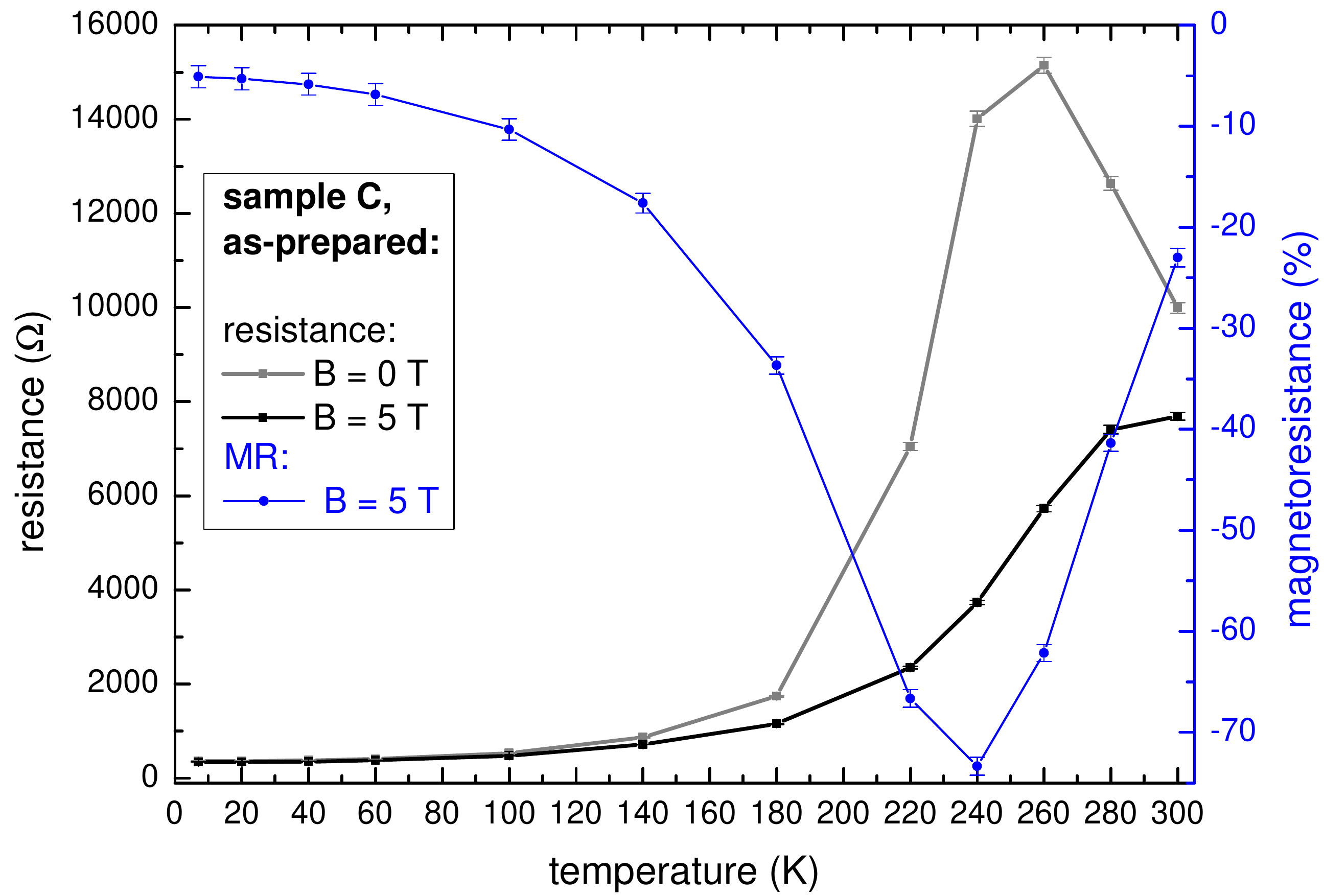}
\caption{\label{fig5} Resistance and magnetoresistance of the \emph{as-prepared 
sample C}. }
\end{figure}

\begin{figure}
\centering
\includegraphics[width=0.75\textwidth]{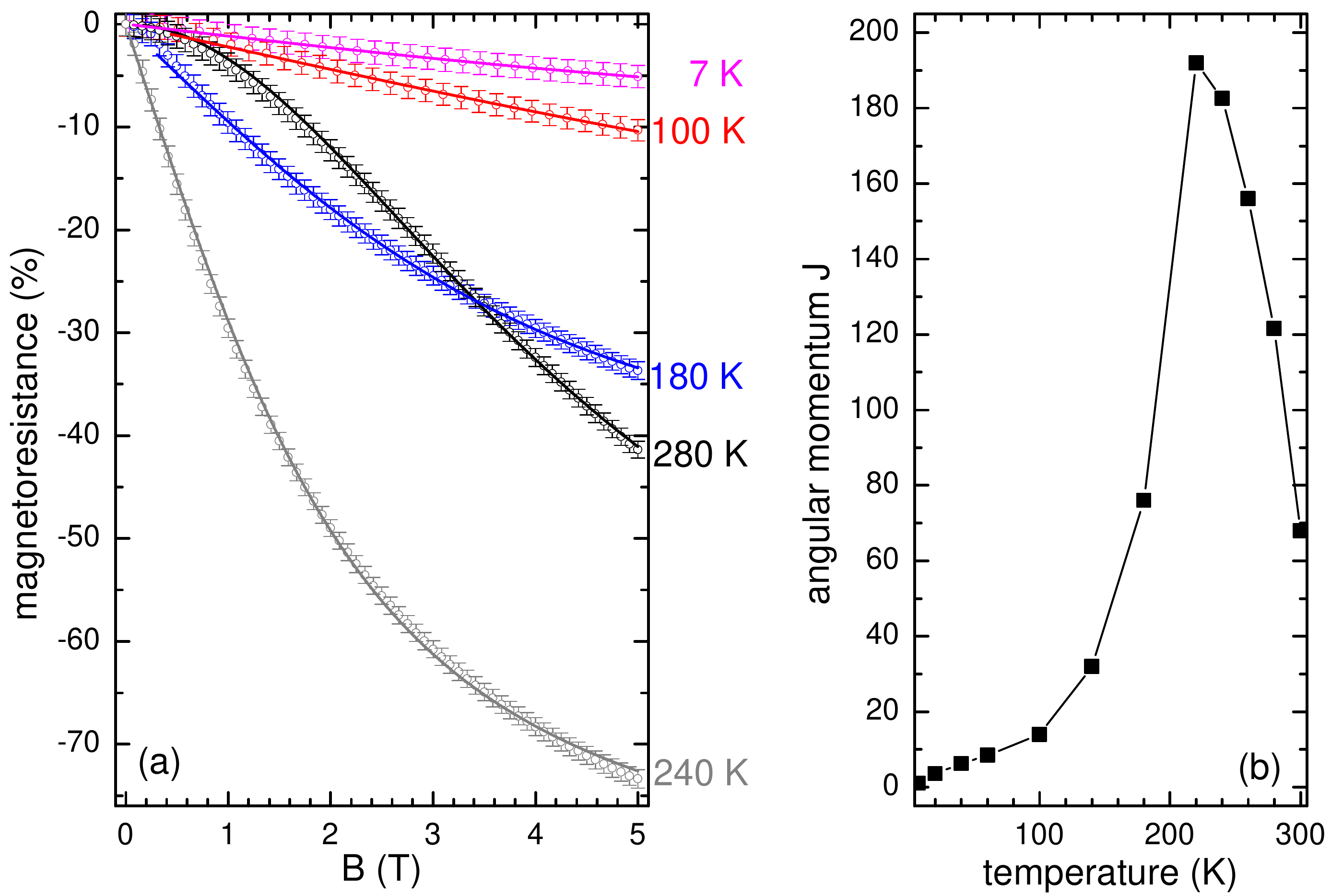}
\caption{\label{fig6} \emph{As-prepared 
sample C}: (a) 
Magnetic-field dependence of the MR for selected temperatures with the
corresponding Brillouin-function fits; (b) angular momentum $J$, as 
derived from the fitting procedures.}
\end{figure}

As a first reference, the magnetoresistance of 
sample C in the non-reduced 
state was studied -- only in the dark, since all non-reduced samples 
did not show any photoconductivity. In a magnetic field of 5~T the resistance 
is clearly reduced and the MIT shifts from 250~K to 300~K (see 
fig.~\ref{fig5}). This means a negative 
magnetoresistance over the whole temperature range from 5 to 300~K, with an extremum 
of -73\% around the MIT. For some selected temperatures, the magnetic-field 
dependence of the MR [fig.~\ref{fig6}(a)] is evaluated by fitting 
the data with the Brillouin function\footnote{For convenience, 
we shortly write $\mathcal{B}$ instead of
$\mathcal{B}_J(g\mu_B J B/k_B T)$ in the following.} 
$\mathcal{B}_J(g\mu_B J B/k_B T)$, see eqs.~(\ref{eq_Brillouin_2}), 
(\ref{eq_Brillouin_1}). 

As expected from theory, the MR above the 
MIT temperature is proportional to $\mathcal{B}^2$ while approaching 
a linear dependence ($\propto \mathcal{B}$) below the MIT. 
$J$ is a free parameter within the 
regression analysis and represents the total angular momentum of the 
magnetic polarons. From fig.~\ref{fig6}(b), which displays $J$ as a function of 
the temperature, we see that $J$ has its maximum in the range of the MIT 
and decreases for both higher and lower temperatures. At 220~K the total angular 
momentum peaks with $J=192$, which corresponds to 96 magnetically ordered Mn$^{3+}$ 
ions. Employing eq.~(\ref{eq_mag_pol_size}), such a magnetic polaron 
has a radius of 1~nm, which is consistent with literature data for other hole-doped 
manganites \cite{wag98,ter97}.

In the metallic phase, the dark resistance between 40$\ldots$110~K is
reproduced by $\rho(T)=\rho_0 + \alpha T^{2.5}$ with 
$\rho_0 =$3.65$\cdot$10$^{-5} \Omega$m, which is a typical behaviour for 
hole-doped mangnites, too \cite{ter96}. However, the value of $\rho_0$ 
is comparatively high for an epitaxial film \cite{gupta_grain-boundary_1996}, 
which is attributed 
to a high defect density due to the nanoscopic CeO$_2$ phase segregation and 
the oxygen excess.   

\subsubsection{Highly reduced sample C}

\begin{figure}
\centering
\includegraphics[width=0.75\textwidth]{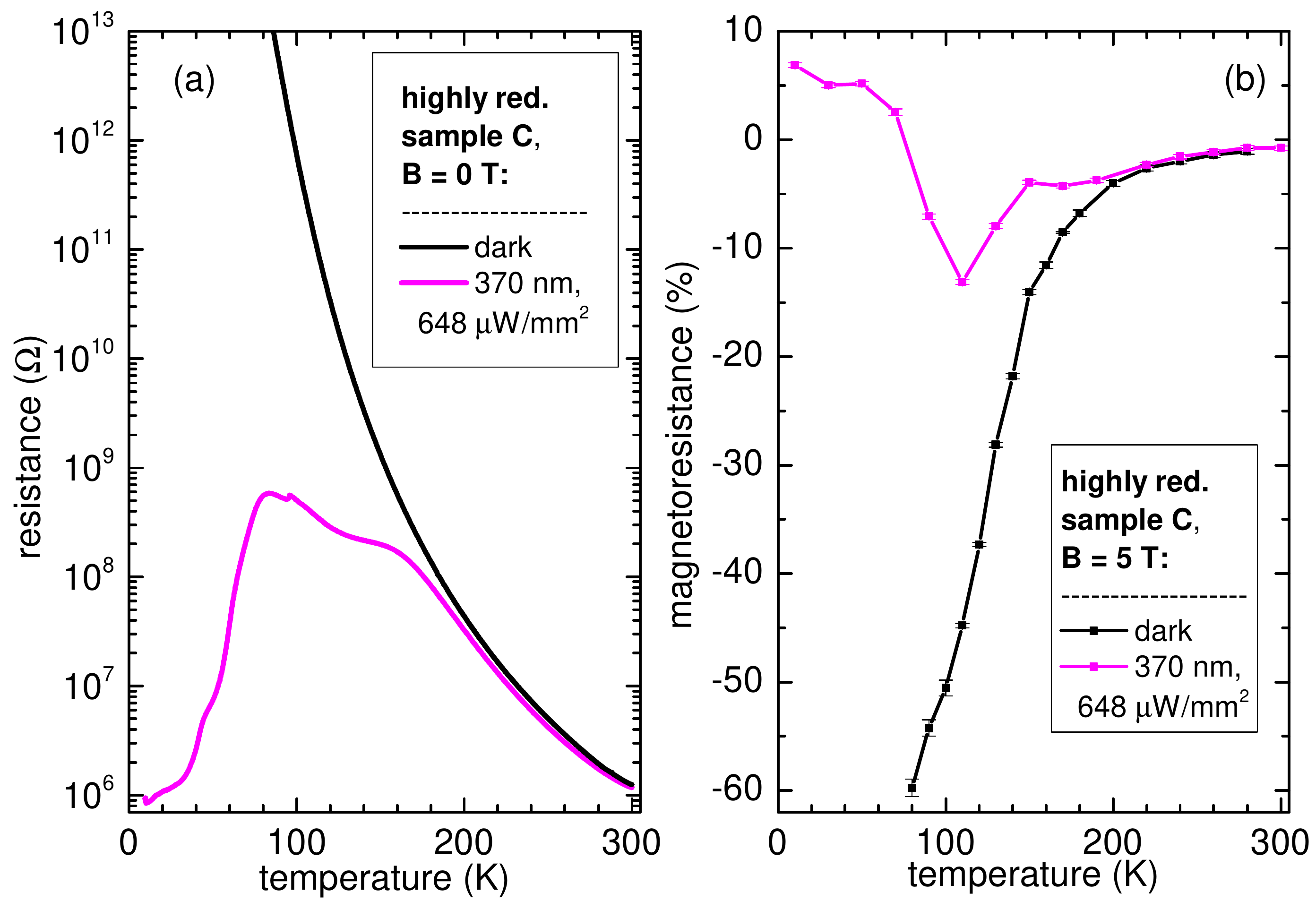}
\caption{\label{fig7} Photoconductivity of the 
\emph{highly reduced sample C}: (a) Resistance vs. 
temperature in zero field in the dark and under 370-nm illumination; 
(b) magnetoresistance in the dark and under illumination.}
\end{figure}

The resistance in zero field and the magnetoresistance at 5~T
of the highly reduced sample C 
in the dark as well as under 370-nm illumination are depicted in 
figs.~\ref{fig7}(a) and (b), respectively. 

\begin{figure}
\centering
\includegraphics[width=0.75\textwidth]{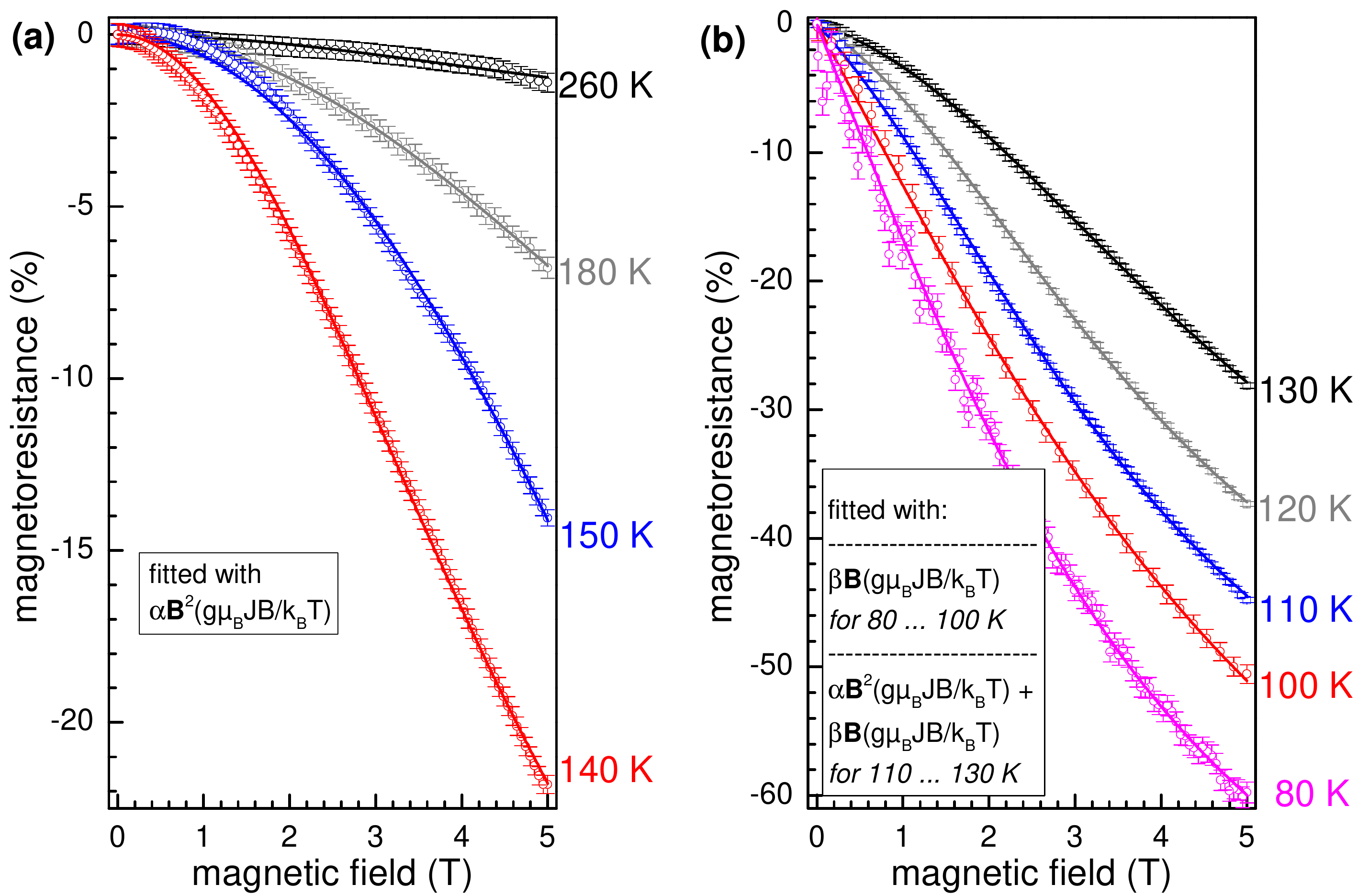}
\caption{\label{fig8} \emph{Highly reduced sample C in the dark}: 
MR-vs.-magnetic-field 
data for selected temperatures, fitted with $\mathcal{B}^2$ for 
higher temperatures (a) and with $\mathcal{B}$ or a linear 
combination of $\mathcal{B}^2$ and $\mathcal{B}$ for lower 
temperatures (b).}
\end{figure}

As found before, \emph{in the dark} the 
sample is insulating in the whole temperature range, while the 
resistance is beyond our detection limit below 80~K. The MR is negative 
and decreases with decreasing temperature down to 
-60\% at 80~K. As shown in fig.~\ref{fig8}(a), the MR vs. B data is reproduced 
by $\mathcal{B}^2$ down to 140~K, with $J$ growing from 20 to 50 with 
decreasing temperature, which corresponds to magnetic-polaron radii between 
0.5~nm and 0.7~nm in accordance with the carrier localization lengths 
of 0.4$\ldots$0.8~nm (as estimated in ref.~\cite{thiessen_hopping_2014}). 
Between 130~K and 110~K 
the MR can be fitted by a linear combination of $\mathcal{B}^2$ 
and $\mathcal{B}$ [fig.~\ref{fig8}(b)], 
while $\mathcal{B}$ reproduces the MR-vs.-T data 
below 100~K [fig.~\ref{fig8}(b)]. A phase transition from the paramagnetic to 
the ferromagnetic state, starting at 130~K, exhibiting a coexistence 
of para- and ferromagnetic areas down to 110~K, and ending in 
a completely ferromagnetic state at 100~K, would be consistent with 
such a behaviour.

\begin{figure}
\centering
\includegraphics[width=0.75\textwidth]{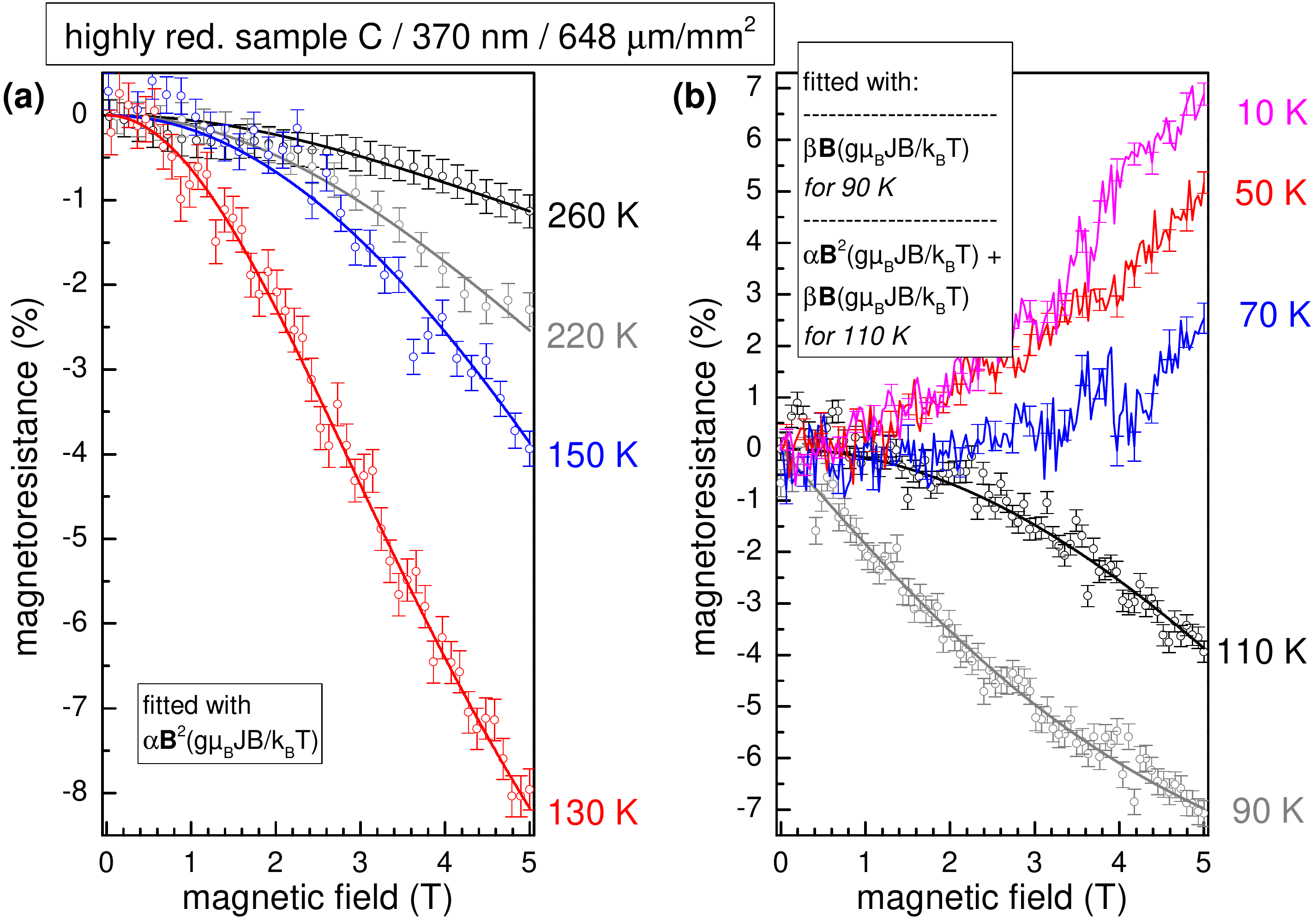}
\caption{\label{fig9} \emph{Highly reduced sample C under 
370-nm illumination}: MR-vs.-magnetic-field 
data for selected temperatures with corrsponding Brillouin-function fits 
down to 90~K. For the three lowest temperatures of 10, 50, and 70~K 
positive MR is observed.}
\end{figure}

\emph{Under illumination}, from 200~K towards lower temperatures, 
the resistance is dramatically lower than in the dark, showing 
a small saturation between 150~K and 80~K [fig.~\ref{fig7}(a)]. 
At 80~K a strong decrease of the 
resistance is visible, while the shape of the R-vs.-T is not typical for 
a mixed-valence manganite having a MIT. The MR at 5~T is negative between 
300~K and 110~K and decreases with falling temperature -- as for the 
dark case. From 110~K towards lower temperatures the MR increases and, 
surprisingly, becomes positive below 70~K, which is non-typical for 
a manganite as well. The magnetic-field dependence of the MR can be 
approximated by $\mathcal{B}^2$ with $J=50\ldots 20$ 
between 300~K and 130~K [fig.~\ref{fig9}(a)], as for the dark case. Also 
similar to the behaviour under dark conditions, at 110~K a linear 
combination of $\mathcal{B}^2$ and 
$\mathcal{B}$ is suitable, while at 90~K 
$\mathcal{B}$ gives the best fit [fig.~\ref{fig9}(b)]. 
However, as a main result, which will be discussed in detail in 
sec.~\ref{sec_discussion_MR}, 
the occurrence of a positive MR is in conflict with a scenario that 
postulates the MIT to stem from
the injection of photogenerated carriers from the substrate into the 
LCeMO film.

\subsubsection{Reduced bare SrTiO$_3$}

\begin{figure}
\centering
\includegraphics[width=0.75\textwidth]{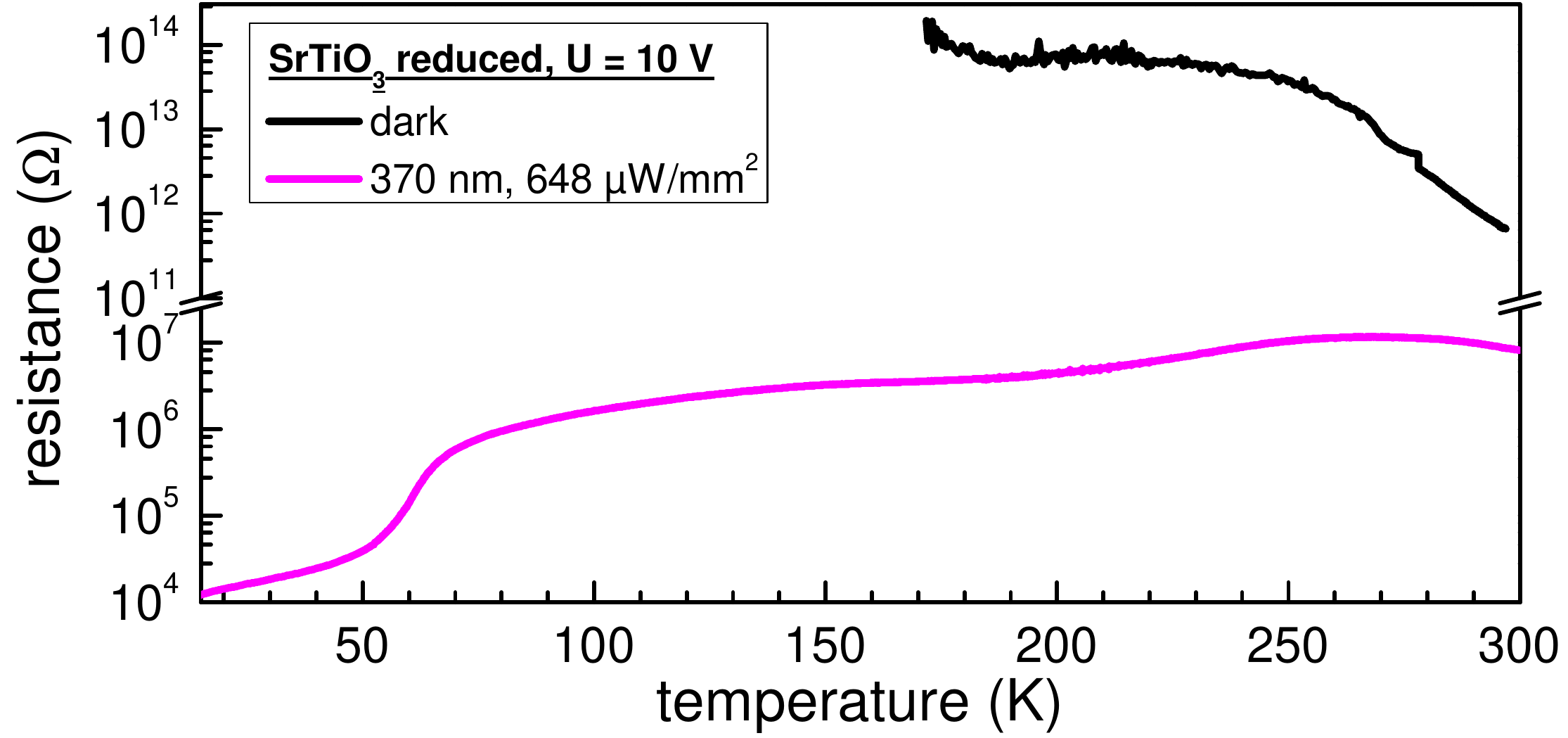}
\caption{\label{figX} Resistance vs. temperature characteristics of the 
bare SrTiO$_3$ reference crystal. Note the break of the resistance axis.}
\end{figure}

\begin{figure}
\centering
\includegraphics[width=0.75\textwidth]{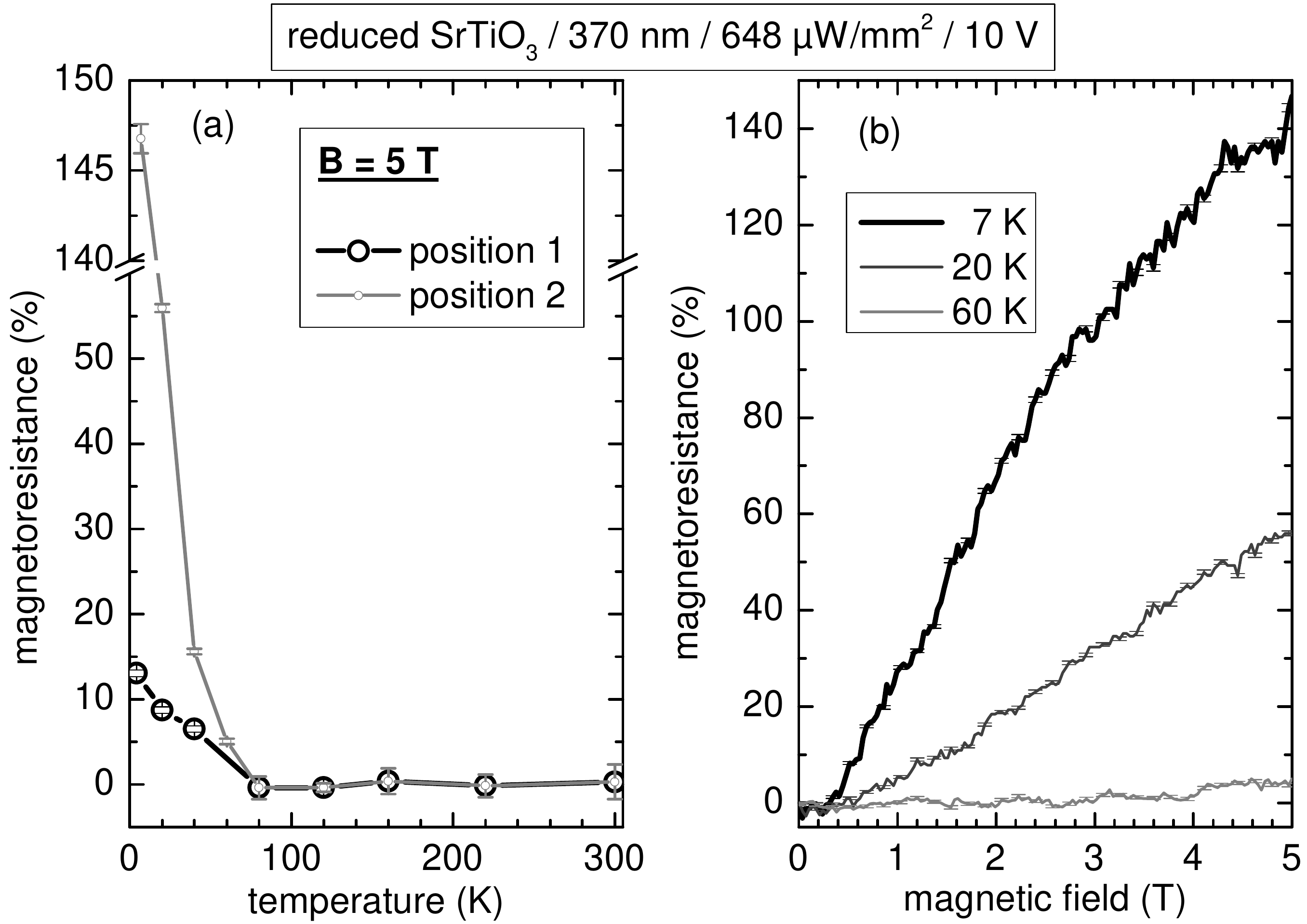}
\caption{\label{figY} Bare SrTiO$_3$ sample: (a) Temperature dependence 
of the MR in a magnetic field of 5~T, measured at two different 
sample surface positions; (b) corresponding magnetic-field dependence of the 
MR at selected temperatures (position 2).}
\end{figure}

As visible in fig.~\ref{figX}, the dark resistance of the reduced SrTiO$_3$ 
can be reliably measured only in the limited range between 300~K and 
250~K. Obviously, the annealing procedure could not increase the electron 
concentration enough to allow transport measurements.Tests with higher 
voltages up to 100~V gave similar results. Illumination with 370~nm at 
648~$\mu$W/mm$^2$ causes a dramatic decrease of the resistance of at least 
five orders of magnitude. We find a maximum of the resistance at 270~K and 
a subsequent slight decrease at lower temperatures, followed by a steeper 
decrease below 70~K [fig.~\ref{fig7}(a)]. Qualitatively, the curve shape is equivalent to the 
respective \emph{R-T} curve of the highly reduced and illuminated sample C for $T<70$~K. 
Furthermore, in 
this low-temperature range, the bare-SrTiO$_3$ resistance is at least 
one order of magnitude smaller than 
the resistance of sample C. Thus, we conclude that the MIT of the highly 
reduced sample C is not primarily 
LCeMO-intrinsic but has its origin in the parallel 
conduction in the photoconductive SrTiO$_3$ substrate. 

The MR of the bare 
SrTiO$_3$ reference was only studied under ilumination due to the very 
high dark resistance. Two data sets with electrical contacts at two different 
locations on the sample surface were recorded, see fig.~\ref{figY}(a) for the 
temperature dependence of the MR at 5~T and fig.~\ref{figY}(b) for the 
magnetic-field dependence of the MR. The resistances at low temperatures 
differ noticably for the two sample locations, which points towards an 
inhomogeneous oxygen diffusion during annealing. Below 60~K a 
positive MR of 5\% 
is observed, at decisively lower temperatures (down to 7~K) values as 
large as 147\% are reached. The MR(B) dependence is approximately linear, 
which was already reported in the 
literature \cite{tufte_magnetoresistance_1968,liu_magnetic-field_2012} 
for highly reduced SrTiO$_3$.

\subsubsection{Discussion of the magnetoresistance results}
\label{sec_discussion_MR}

Here, we consider the implications of the MR results on our understanding 
of the photoconductivity results, while the implications on the 
interpretation of the electroresistive effect of sample D have been 
exported to the supplement.

Regarding the reference MR measurements on the bare reduced SrTiO$_3$ 
single crystal, both the light-induced MIT at 80~K and the positive MR 
below 70~K in the highly reduced 
sample C clearly reflect the occurrence of parallel charge transport within 
the SrTiO$_3$ substrate. However, the plateau in the \emph{R-T} curves of the 
highly reduced samples B and C under 370-nm illumination might still be 
interpreted as an intrinsic effect from the manganite films, i.e., the 
electron concentration is increased by photogenerated-charge injection 
from the substrate and/or by photogenerated carriers from the film itself, 
but not strong enough to cause a complete delocalization of the 
carriers \cite{seeger_charge_1999}. Thus, the photoconductivity above 80~K 
is in agreement with the XPS results of ref.~\cite{thiessen_mn2+/mn3+_2014}. 
However, one discrepancy between the former XPS work and the present 
results remains: In the latter, no photoconductivity at room temperature 
was observed, while in the first, there is a clear change of the Mn 
valence when illuminating at 300~K. One possible explanation hints towards 
photoinduced electron generation confined to the sample surface, 
for which the XPS is sensitive. The effect might be too small to decisively 
change the electron concentration in the whole film, which is the 
reason, why the present transport measurements cannot detect this effect. 
Finally, from all methods used here and previously (resistance 
measurements, XPS, magnetoresistance -- all in the dark and under illumination), 
one can summarize that none of the contributions, which were discussed 
to contribute to the observed photoconductivity, i. e., (i) intrinsic 
carrier generation in the LCeMO films, (ii) injection of carriers from the 
photoconductive SrTiO$_3$ substrate into the film, and (iii) parallel 
(photo-)conduction in the SrTiO$_3$, can be totally excluded. All three 
effects may coexist, however with different weights depending on the 
temperature and the illumination parameters.

\section{Summary and conclusions}

Differently thick 
La$_{0.7}$Ce$_{0.3}$MnO$_3$ films on SrTiO$_3$ with a variable degree 
of oxygen content and CeO$_2$ phase segregation were studied with respect 
to its conductivity and magnetoresistance under illumination, in order to 
achieve a deeper understanding of the photoconductivity and the light-induced 
metal-insulator transition.

While formerly the coexistence of (i) photogenerated carriers within the the LCeMO films  
and (ii) carriers injected from the substrate or from interface states was proposed 
to be the origin for the \emph{R-T} characteristics under illumination, the 
observation of \emph{positive} magnetoresistance under illumination for both 
the LCeMO/SrTiO$_3$ heterostructures and a bare SrTiO$_3$ single crystal strongly 
points towards a third (coexisting) mechanism influencing the measured resistance 
behaviour: parallel photoconduction in the substrate.

Unexpectedly, the highly reduced LCeMO film with the strongest CeO$_2$ phase 
segregation showed an electrical-field induced MIT, which could, 
however, after the evaluation of the magnetoresistance behaviour, be 
ascribed to the charge transport in the SrTiO$_3$ substrate, too.

Finally, we conclude that the role of the 
SrTiO$_3$ substrate in previous studies of the photoinduced properties 
of manganite thin films, e.g., 
refs.~\cite{bey09,bey10,gil00,cau01}, has to be questioned 
critically. In the future, it might be more reasonable to perform 
investigations of the intrinsic photoconductivity of 
manganites on either films grown on less photoresponsive 
substrates than SrTiO$_3$ (e.g., LaAlO$_3$) or on bulk crystals.


\section*{Acknowledgement}

This work was kindly supported by the German Science 
Foundation (DFG, grant no. BE 3804/2-1).

\section*{References}



\end{document}